\documentclass[a4paper]{article}
\usepackage{INTERSPEECH2018}

\title{CNN+LSTM Architecture for Speech Emotion Recognition with Data Augmentation}

\name{Caroline Etienne $^{1,2,*}$, Guillaume Fidanza $^{2,*}$, Andrei Petrovskii $^{2,*}$,\\ Laurence Devillers $^1$, Benoit Schmauch $^2$.}
  
\address{
  $^1$ LIMSI, CNRS, Paris-Sud University, Paris-Saclay University / F-91405 Orsay, France\\
  $^2$ DreamQuark, 29 rue de Courcelles, 75008 Paris, France\\
  $^*$ Equal contribution}
\email{caroline.etienne@limsi.fr, guillaume.fidanza@dreamquark.com, andrei.petrovskii@dreamquark.com, laurence.devillers@limsi.fr, benoit.schmauch@dreamquark.com}
\date{}

\usepackage{times}
\usepackage{latexsym}

\usepackage{url}

\usepackage{microtype}
\usepackage{graphicx}
\usepackage{subfig}
\usepackage{booktabs}
\usepackage{hyperref}
\usepackage{hyperref}
\usepackage{multirow}
\usepackage{amsmath}
\usepackage{subfiles}

\newcommand{\beq}{\begin{equation}}
\newcommand{\eeq}{\end{equation}}
\newcommand{\bs}{\boldsymbol}  

\usepackage[table]{xcolor}
\definecolor{mygray}{RGB}{238,238,238}
\definecolor{myblue}{RGB}{176,224,230}

\begin{document}
\maketitle
\begin{abstract}
In this work we design a neural network for recognizing emotions in speech, using the IEMOCAP dataset. 
Following the latest advances in audio analysis, we use an architecture involving both convolutional layers, for extracting high-level features 
from raw spectrograms, and recurrent ones for aggregating long-term dependencies. We examine the techniques of data augmentation with vocal track length perturbation, 
layer-wise optimizer adjustment, batch normalization of recurrent layers and obtain highly competitive results of $64.5\%$ for weighted accuracy and $61.7\%$ for
unweighted accuracy on four emotions.
\end{abstract}

\section{Introduction}
Providing high quality interaction between a human and a machine is a very challenging and active field of research with numerous applications. 
An important part of this domain is recognition of human speech emotions by computer systems. In the last years, impressive progress has been achieved in 
speech recognition by means of deep learning \cite{Baidu,Medennikov+2016,Saon+2016,gatedCNN}. These achievements also include significant results on speech 
emotion recognition (SER), see e.g. \cite{KimLP13,microsoft,tspredictor}. 
In this work we build a neural network for SER on the IEMOCAP dataset \cite{Busso2008} and achieve the result highly competitive to the state of the art. 
\footnote{To our knowledge, the present state of the art has been achieved in \cite{microsoft}. However the cross-validation procedure performed in this 
paper (as in other works presenting the results obtained on the IEMOCAP dataset) includes only five folds of the dataset out of the ten 
possible. On the other hand, our experiments showed (see section \ref{modexp}) that the performance strongly depends on the part of the data which is used 
for measuring the scores. As a consequence the results obtained by 5-fold cross-validation without clarification what data has been used for the measurement 
are not possible to compare with. 
Therefore we propose to use 10-fold cross validation as the correct way for measuring the scores on IEMOCAP dataset and present our results correspondingly.} 

When treating a SER problem with deep learning, one either creates hand-crafted acoustic features (MFCC, pitch, energy, ZCR...), which are used as inputs to a 
neural network, or sends the data, after some preprocessing (e.g. Fourier transform), directly to a neural network. We apply the second strategy by transforming 
the audio signal to a spectrogram, which is then used as an input to convolutional layers, followed by recurrent ones. Such a choice of an architecture, 
which has recently demonstrated very competitive performance \cite{clstm,Baidu,tspredictor}, assumes two main interpretations. On one hand, adding few convolutional 
layers in the beginning of the network is an efficient way to reduce dimensionality of the data and can significantly simplify the training procedure. On the other hand, 
it is also possible to use a deep CNN for extracting high-level features, which are then fed to a RNN for final time aggregation. We test a variety of architectures 
with different depths for the convolutional (1-6 layers) and recurrent modules (1-4 Bi-LSTM layers), achieving the best scores with a 4+1 scenario\footnote{4 convolutional 
and 1 Bi-LSTM layers}. 

To address challenges of class imbalance and data scarcity, we explored a vocal tract length perturbation for the purpose
of data augmentation, and showed that it significantly improves the performance.
In line with \cite{BNrecNN,Baidu,RBN,LN} we examined batch normalization applied to the recurrent layers of the network. Finally, we noticed that parameters of convolutional and Bi-LSTM layers are trained at a very different pace. 
We tried to take advantage of this observation by per-layer adjustment of the update rule parameters, but unfortunately were not able to make a definite conclusion in favor of this idea.

\subsection{Dataset description}
\label{datadescription}
IEMOCAP (Interactive Emotional Dyadic Motion Capture), collected at the University of Southern California (USC) \cite{Busso2008}, is
one of the standard datasets for emotion recognition. It consists of twelve hours of audio and video recordings performed by 10 professional actors (five women and five men) and organized in 5 sessions of dialogues between two actors of different genders, either playing a script or improvising. 
Each sample of the audio set is an utterance assigned with an emotion label. Labeling was made by six students of USC, three at a time for each utterance. 
The annotators were allowed to assign multiple labels if necessary. The final true label for each utterance was chosen by majority vote if the emotion 
category with the highest vote was unique. 
Since the annotators reached consensus more often when labeling improvised utterances (83.1\%) than scripted 
ones (66.9\%) \cite{Busso2008}, we concentrate only on the improvised part of the dataset. For the sake of comparison with the prior state-of-the-art approaches, 
we predict four of the most represented emotions: neutral, sadness, anger and happiness, which leave us 2280 utterances in total.

\section{Data augmentation}

\begin{figure}
     \centering
     \subfloat{\includegraphics[scale=0.45]{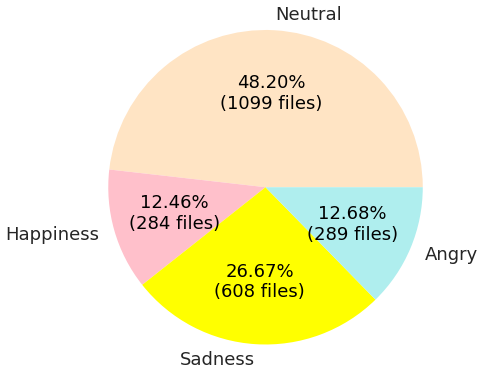}}
     \caption{Class distribution of the utterances in the improvised part of the IEMOCAP dataset}
     \label{fig:class-distribution}
\end{figure}

The IEMOCAP dataset has two main drawbacks: class imbalance (see Fig. \ref{fig:class-distribution}) and small size. 
To cope with both obstacles, we examined data augmentation by means of vocal tract length perturbation (VTLP), at the same time oversampling the least represented classes of 
the dataset: happiness and anger. VTLP is based on the speaker normalization technique considered in \cite{FWN}, where it was implemented to reduce interspeaker variability. The difference in human's vocal tract 
length can be modeled by rescaling the peaks of significant formants along the frequency axis with a factor $\alpha$ taking values in the approximate range $(0.9, 1.1)$.
Therefore, in order to get rid of this variablility, one should estimate the factor for each speaker and accordingly normalize the spectrograms. Applied inversely, the same idea 
can be used for data augmentation \cite{Jaitly2013VocalTL,IBMAUG, yerevan2016}: in order to generate new samples, one simply has to perform rescaling of the original spectrograms 
along the frequency axis while keeping the scaling factor in the range $(0.9, 1.1)$. Both approaches, normalization and augmentation, pursue the same objective: to enforce the 
invariance of the model to speaker-dependent features, since they are not relevant to the classification criterion. Augmentation, however, is easier to implement because we don't 
need to estimate the scaling factor of each speaker, and therefore we stick to this option.

Rescaling of frequencies has been performed as follows \cite{FWN}:
\beq
G(f) =
\begin{cases}
   \alpha f & 0\leq f\leq f_0 \\
   \frac{f_{max} - \alpha f_0}{f_{max} - f_0}(f-f_0) + \alpha f_0 & f_0\leq f\leq f_{max},
 \end{cases}
\eeq
where $f_{max}$ is the upper cut-off frequency and $f_0$ is defined to be larger than the highest significant formants (we took $\frac{f_0}{f_{max}}=0.9$). 
Therefore, we rescale the frequencies below $f_0$ with $\alpha\in (0.9, 1.1)$, and then rescale the rest to ensure that the considered diapason stays constant. 

We tried two strategies of data augmentation. In the first one, a single uniformly distributed value $\alpha\in (0.9, 1.1)$ was sampled at each epoch and used to 
rescale all training examples, and no rescaling was applied to the validation set. In the second strategy, each spectrogram was rescaled with an individually 
generated $\alpha$ for the training, as well as for the validation sets. For evaluation, we used the majority vote of the model predictions on eleven copies of 
the test set with $\alpha = 0.9, 0.92, 0.94,...,1.1$. We present the scores obtained with the second augmentation strategy, which provided the best result.

\section{Model description and experiments}
\label{modexp}
As it has been mentioned above, the IEMOCAP dataset consists of five sessions, each being a conversation between a man and a woman, giving 10 speakers in total. 
In order to see how well the model can generalize to different speakers, we took the validation and test sets to correspond to two different speakers of one of the sessions. 
The training set was composed of the four remaining sessions.
In the course of experiments, we observed that the performance strongly depends on which speakers are chosen for the test set (see Tab. \ref{table-folds}). 
Therefore we choose 10-fold cross-validation strategy, in order to average over all possible choices of the dataset splitting. Interestingly, to the best of our knowledge, 
all the other results reported on the IEMOCAP dataset were obtained by 5-fold cross-validation. In this case the choice of the validation and test sets is not 
rigorously defined\footnote{For instance, one could systematically use female speakers as validation and male speakers as test, or inversely} and the scores 
obtained in this way are not possible to compare with.

For evaluating the model performance, we chose weighted (WA) and unweighted (UA) accuracies. WA is the standard accuracy computed over the whole test set. UA is an average over 
accuracies computed for each emotion separately. First, we compute the metrics for each fold and then present the scores as the average over all the folds.
Since for imbalanced datasets UA is a more relevant characteristic, we rather concentrated our efforts on getting a high UA, in line with most of the other works on IEMOCAP.

\begin{figure*}
     \centering
     \subfloat{\includegraphics[scale=0.38]{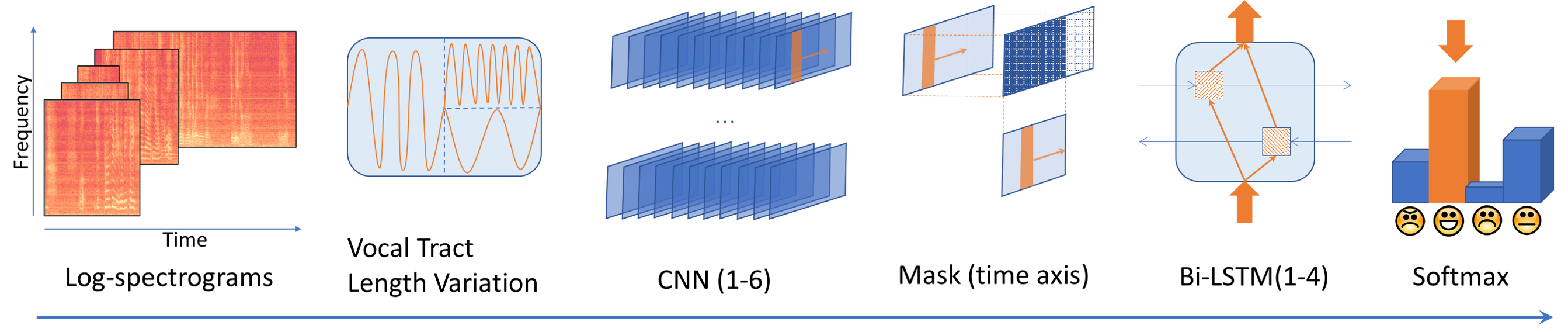}}
     \caption{Network architecture}
     \label{fig:archi}
\end{figure*}

We considered architectures with 1-6 convolutional layers, 1-4 Bi-LSTM layers and a dense layer with softmax nonlinearity on top of the network (see Fig. \ref{fig:archi}). 
As an optimization procedure, we used stochastic gradient descent with momentum and the batch size of 16\footnote{We chose the small batch size in order to achieve high variability in the gradient descent directions}. For the regularization of weights we used L2-regularization.

Due to the significant variety of the data samples in the time length (from 21 to 909 time steps for window size $N=64ms$ and shift $S=32ms$) we performed zero-padding of 
the samples along the time axis. In order to avoid the aggregation of the artificially added time steps by Bi-LSTM, we put a masking layer between the 
convolutional and Bi-LSTM modules. The size of the mask has been derived from the temporal size of the corresponding spectrogram and action of the convolutional strides on it.   

Finally we normalized the samples according to the general statistics of the dataset:
\beq
x^n = \frac{x-\hat{x}}{\sqrt{\sigma^2+\epsilon}},
\eeq
where $\hat{x}$ and $\sigma$ are the average and standard deviation of the spectrogram pixels computed over the whole dataset along both time and frequency axes. 
Such normalization significantly improves the convergence time of the model. However, applied to networks of small depth ($\leq2$ convolutional layers), 
it results in strong overfitting.

As we have mentioned above, we conducted a variety of experiments with different depths of convolutional and Bi-LSTM modules. The presence of pooling layers alternating 
with the convolutions noticeably decreased the performance and has been discarded in the beginning of the experiments. We examined different scenarios: 
"shallow CNN + deep Bi-LSTM", "deep CNN + shallow Bi-LSTM" and "deep CNN + deep Bi-LSTM", mostly concentrating on the second option. The best results has been achieved with a choice of 4 convolutional and 1 Bi-LSTM layers. 

In Tab. \ref{table-comparison} we present the results of the best model and also contributions to the performance of the techniques we applied. One can see that 
oversampling allowed to increase UA by $2.1\%$, but resulted in $2.9\%$ decrease of WA. Data augmentation with VTLP led to increase of both metrics by $1.1\%$ and $0.7\%$ 
for UA and WA correspondingly. Considering a larger range of the frequencies (8kHz) increased the UA by $0.8\%$. Finaly in Tab. \ref{table-folds} we present the results per fold, the scores obtained by averaging our 5 best folds and the results obtained in the other works by 5 fold cross-validation.

We also tried out batch normalization implemented for the Bi-LSTM layers of the network. During the experiments we observered that the data of interest are sensitive 
to normalization. Therefore we choose the most conservative normalizing strategy which implies averaging the samples over all the axes:
\beq
\pi_{s,t,f}^n = \frac{\pi_{s,t,f}-\hat{\pi}}{\sqrt{\sigma^2+\epsilon}},
\eeq
where
\beq
\hat{\pi} = \frac{1}{b_{tf}}\sum\limits_{s,t,f}{\pi_{s,t,f}}, \quad \sigma = \frac{1}{b_{tf}}\sum\limits_{s,t,f}{(\pi_{s,t,f}-\hat{\pi})^2}.
\eeq
Here, $s$, $t$ and $f$ are the batch, temporal and frequency index respectively, $\pi$ is preactivation
and $b_{tf}$ is a product of the sum of the sample time lengths over the batch and the feature number.
Then batch normalization is applied only to the input contribution to the hidden state:
\beq
h_t = a(W_h h_{t-1}+BN(W_x x_t)),
\label{bn_seq}
\eeq
where $BN$ stands for the standard batch normalization operation \cite{DBLP:journals/corr/IoffeS15}, $a(\pi)$, $h_t$, $x_t$ are activation, hidden state and input, 
and $W_h$, $W_x$ are the corresponding weights.

\begin{table*}[!htb]
\caption{10-cross validation scores depending on the techniques applied (for each experiment we present the results corresponding to its best run).}
\label{table-comparison}
\centering
\begin{tabular}{lccccc >{\columncolor{mygray}}c}
\toprule
 & Baseline & & & Best model\\
\midrule
Augmentation during training & - & - & + & +\\
Oversampling ($\times 2$) of happiness and anger & - & + & + & +\\
Frequency range (kHz) & 4 & 4 & 4 & 8\\
\midrule
Weighted accuracy & 66.4 & 63.5 & 64.2 & 64.5\\
Unweighted accuracy & 57.7 & 59.8 & 60.9 & \textbf{61.7}\\
\bottomrule
\end{tabular}
\end{table*}

\begin{table}[!htb]
\caption{The performance of the best model per fold and comparison to the other works. The gender column indicates which speaker is used as test set in the fold.}
\label{table-folds}
\centering
\begin{tabular}{ccccc}
\toprule
Fold & Session   & Gender & WA (\%) & UA (\%)\\
\midrule
1 & 1 & F & 64.1 & 66.4\\
\rowcolor{mygray} 2 & 1 & M & 68.8 & 67.7\\
3 & 2 & F & 70.3 & 71.3\\
\rowcolor{mygray} 4 & 2 & M & 62 & 67.6\\
5 & 3 & F & 64.8 & 52.1\\
\rowcolor{mygray} 6 & 3 & M & 66.4 & 56\\
7 & 4 & F & 68.5 & 59.7\\
\rowcolor{mygray} 8 & 4 & M & 64.3 & 67.3\\
9 & 5 & F & 64.8 & 64.2\\
\rowcolor{mygray} 10 & 5 & M & 51 & 44.2\\
\multicolumn{3}{c}{10 fold cross-valid.} & \textbf{64.5} & \textbf{61.7}\\
\multicolumn{3}{c}{5 best folds} & \textbf{66.9} & \textbf{65.3}\\
\multicolumn{3}{c}{\cite{microsoft} (5 fold cross-valid.)} & 62.9 & 63.9\\
\multicolumn{3}{c}{\cite{tspredictor} (5 fold cross-valid.)} & 67.3 & 62.0\\
\bottomrule
\end{tabular}
\end{table} 

We examined batch normalization applied to the architectures with 4 convolutional and 1-4 Bi-LSTM layers. 
The experiments with the initial batch size of 16 demonstrated faster overfitting and degradation of the performance compared to the baseline. 
The further experiments with larger batch size showed that it strongly influences the performance (see Tab. \ref{table-batchnorm}), despite the fact that the normalization 
has been performed along all the axes of the batch. One can see that the scores obtained with the batch size of 64 almost reaches the performance of our best model \ref{table-comparison}.  
Therefore, it is possible that further augmenting the batch size would lead to even better results. Unfortunately, due to GPU memory restrictions, we could not verify it. 

\begin{table}[!htb]
\caption{The performance of the best model equipped with the batch normalization (for each experiment we present the results corresponding to its best run).}
\label{table-batchnorm}
\centering
\begin{tabular}{lccc}
\toprule
Minibatch size & 16 & 32 & 64\\
\midrule
Weighted & 63.6 & 65.1 & 65.4\\
Unweighted & 58.9 & 59 & 60.8\\
\bottomrule
\end{tabular}
\end{table}

\subsection{Difference in the gradient scaling of the convolutional and recurrent layers}
\label{lwga}

\begin{figure}
     \centering
     \subfloat{\includegraphics[scale=0.35]{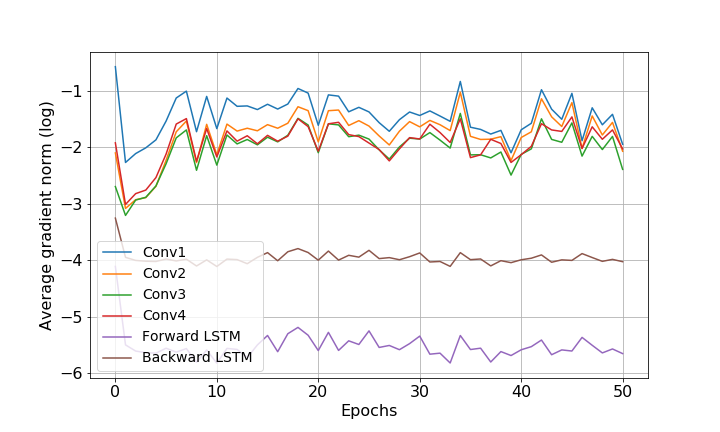}}
     \caption{Per-layer gradient evolution}
     \label{fig:grad_evol}
\end{figure}

Monitoring the gradient of the network parameters, we observed that the gradient with respect to the weights of the convolutional layers is much larger than with respect to the weights of Bi-LSTM (see Fig. \ref{fig:grad_evol}). This observation allows an interpretation that regarding the convolutional weights the loss surface should be steeper and deeper than regarding the weights of the Bi-LSTM. Therefore it gave us a nudge that it might be interesting to consider different update parameters, namely learning rate and momentum, for convolutional and recurrent modules. 
Apart from varying the conventional update parameters of the momentum optimizer we also considered its modification by introducing the new parameter $\beta$: 

\beq
\bs{v}_t = \bs{v}_{t-1}\gamma + \eta \nabla_{\bs{w}} J(\bs{w}), \quad
\bs{w} = \bs{w} - \beta \bs{v}_t,
\label{modopt}
\eeq
where $\gamma$, $\eta$ and $\bs{v}_t$ stands for the momentum, learning rate and velosity correspondingly. The coefficient $\beta$ brought into use in this way does not accumulate in the velocity expression and provide better control of the momentum term of the optimizer.  
Unfortunately, from our experiments we were not able to draw any definite conclusion in favor of layer-wise adjustment of $\eta$, $\gamma$ or $\beta$. Nevertheless, we find that this is an interesting direction to persue and more thorough experiments might give more preferable result. 

It also might be interesting to test the update rule modification introduced in eq. (\ref{modopt})
in the other settings in order to see whether it can provide an actual improvement of the momentum optimizer.

\section{Conclusion}

In this work we built a neural network for recognizing emotions in speech, using the IEMOCAP dataset. Unlike the prior results, in order to measure the model performance 
we performed 10-fold cross-validation, which is more appropriate for this dataset.
To adress the issues of scarcity and class imbalance we employed data augmentation by means of VTLP and minor class oversampling. 

Following the modern trends in speech analysis, we used a mixed CNN-LSTM architecture, exploiting the capacity of convolutional layers to extract high-level representations 
from raw inputs. Interestingly, we noticed that parameters of convolutional and LSTM layers are trained at a very different pace. We tried to take advantage of this observation by per-layer adjustment of the update rule parameters, but unfortunately were not able to make a definite conclusion in favor of this idea. Nevertheless, we find that this is an interesting direction to persue and more thorough experiments might give more preferable result. 

We also investigated the effect of batch normalization, an indispensable tool in most image recognition tasks. In order to preserve the signal structure as much as possible 
we performed the normalization layer-wise as well as batch-wise. Nevertheless, we did not manage to increase performance compared to the baseline, which might be 
caused by the small batch size we had to use in order to fit into the available GPU memory.

\bibliography{biblio}
\bibliographystyle{IEEEtran}

\end{document}